\documentclass[12pt]{article}
\usepackage{amsfonts,amsmath,amssymb}
\usepackage[pdftex]{graphicx}
\usepackage{hyperref}
\usepackage{float}

\newcommand{\Ord}{{\cal O}}

\makeatletter

\@addtoreset{equation}{section}
\makeatother

\def\href#1#2{#2}

\begin{document}

\begin{titlepage}

\begin{center}

\hfill 
\vskip18mm

\textbf{Localized Towers on the Composite Magnetic Monopole and\\[1mm] 
the Weak Gravity Conjecture$=$Distance Conjecture Connection\\[1mm]
in Effective Field Theories}\\[12mm]

\renewcommand{\thefootnote}{\fnsymbol{footnote}} 
\setcounter{footnote}{1}

{Moreshwar Pathak
and 
Kazuyuki Furuuchi\footnote{Corresponding author.}}\\[6mm]

{\sl Manipal Centre for Natural Sciences,}\\
{\sl Manipal Academy of Higher Education,}\\
{\sl Manipal 576 104, Karnataka, India}\\[3mm]
\end{center}

\vskip8mm 

\begin{abstract}
We study the energy spectrum of the composite magnetic monopole (CMM)
in the model originally constructed in \cite{Saraswat:2016eaz}
to examine how the applicability of the weak gravity conjecture (WGC)
propagates to lower energy scales in the effective field theory (EFT) framework.
This is motivated by 
the WGC$=$distance conjecture (DC) connection
suggested in the literature.
Assuming that the limits of the model parameters
correspond to the boundaries of the moduli space in a UV theory,
we showed that the parameter dependence
of the energy spectrum of the CMM
is broadly in accord with the prediction
of the WGC$=$DC connection.
However, 
unlike the original DC,
the low-energy excitations of the CMM are localized on the CMM.
This leads us to propose an extension of the DC to
include the localized tower of excited states.
We discuss the implications of this extension 
of the DC to the swampland constraints on EFTs.
\end{abstract}

\end{titlepage}

\renewcommand{\thefootnote}{\arabic{footnote}} 
\setcounter{footnote}{0}

\section{Introduction}\label{sec:intro}

Swampland 
conjectures are proposed criteria for
identifying
effective field theories (EFTs)
that are not consistently coupled to quantum gravity
\cite{Vafa:2005ui}.
The weak gravity conjecture (WGC)
\cite{Arkani-Hamed:2006emk} is one of the most extensively studied swampland conjectures.
The WGC proposes that for an EFT with an Abelian gauge symmetry 
to be consistent with quantum gravity,
there must exist a state whose charge $e$ and mass $m$ satisfy
(electric weak gravity conjecture, eWGC)
\begin{equation}
e \gtrsim \frac{m}{M_P} \,,
\label{eq:eWGC}
\end{equation}
where $M_P:= 1/\sqrt{8 \pi G}$ is the reduced Planck mass
($G$ is the Newton's gravitational constant).
We obtain an interesting bound on the UV cut-off scale $\Lambda_{\rm UV}$ of the EFT
if we apply the WGC to magnetic charges (mWGC):
The magnetic charge 
of the unit charge Dirac monopole
is given as $2\pi/e$ (the Dirac quantization condition),
while its mass $m_{\rm mono}$ is estimated as\footnote{%
\eqref{eq:monopolemass} is actually the lower bound:
If the mass of the magnetic monopole is
determined by the physics at $\Lambda_{\rm higher} > \Lambda_{\rm UV}$,
we have a bound tighter than \eqref{eq:mWGCbound}: 
$\Lambda_{\rm UV} < \Lambda_{\rm higher} \lesssim e M_P$.
We will encounter such an example in Sec.~\ref{sec:CMM}.
In the meantime, we will use the estimate \eqref{eq:monopolemass} unless stated otherwise.}
\begin{equation}
m_{\rm mono} \sim \frac{\Lambda_{\rm UV}}{e^2}  \,. 
\label{eq:monopolemass}
\end{equation}
It follows from the mWGC that
\begin{equation}
\Lambda_{\rm UV} \lesssim e M_P \,.
\label{eq:mWGCbound}
\end{equation}
In string theory,\footnote{We regard string theory 
as the unifying theory of all forces in Nature, consistently including quantum gravity.}
a gauge coupling $e$ in the low-energy EFT is often related to 
the vacuum expectation value of some canonically normalized scalar field $\phi$ as
\begin{equation}
e \sim e_0\, \exp \left[ - \alpha \frac{\phi}{M_P} \right]\,,
\label{eq:e}
\end{equation}
where $e_0$ is a constant, and $\alpha$ is a constant of order one.
The scalar field can be the dilaton or the volume moduli, 
for example.
This exponential dependence
of coupling constants
on scalar fields
is suggestive of the connection to another
well-studied swampland conjecture
called the distance conjecture (DC)  \cite{Ooguri:2006in}:
The DC proposes that 
as a scalar field $\phi$ travels trans-Planckian scales
(we will express this situation as 
``a scalar field approaches a boundary of the moduli space''\footnote{%
``Infinite distance limit'' and ``a tower of massless states'' 
are popular expressions in the literature describing this situation.
In string theory, we may take such limits 
(unless another dual frame is required),
for example,
the emergent string conjecture (ESC) \cite{Lee:2019wij,Lee:2019xtm}
proposes that the lightest tower in the DC is either 
weakly coupled string excitations or Kaluza-Klein (KK) modes,
both can be described by string theory.
However, the current work
focuses on EFT descriptions.
In this case, the EFT descriptions 
are expected to break down 
before reaching the infinite distance:
In fact, in Planck units that are natural for discussing swampland conjectures,
the EFT descriptions would break down when the scalar field value exceeds 
just a little bit above one.
Therefore, we use the term ``boundary,''
indicating that beyond the boundary 
the EFT descriptions would cease to be applicable.}),
a tower of light states appears, whose mass $m$ decreases as
\begin{equation}
m \sim m_0 \exp \left[ -\alpha' \frac{\phi}{M_P} \right] \,.
\label{eq:DC}
\end{equation}
Here, $\alpha'$ is a constant of order one with $\alpha' \phi > 0$, and 
$m_0$ is a constant mass scale. 
Comparing \eqref{eq:mWGCbound} with \eqref{eq:e}
and \eqref{eq:DC},
it is tempting to 
identify the energy scale $eM_P$ bounding the EFT UV cut-off scale in the mWGC \eqref{eq:mWGCbound}
and the mass scale of the tower of light states
appearing in the DC \eqref{eq:DC} (mWGC$=$DC conjecture)
\cite{Klaewer:2016kiy}.
In other words,
the mWGC$=$DC conjecture proposes that
above $\Lambda_{\rm UV}$,
a tower of states whose mass scale is given by $\Lambda_{\rm UV}$ appears,
and the EFT description breaks down.

Some stronger versions of the WGC have been proposed
\cite{Heidenreich:2015nta,Montero:2016tif,Heidenreich:2016aqi,Andriolo:2018lvp},
which state that a tower of charged states must satisfy the WGC \eqref{eq:eWGC}
in order not to be in the swampland.\footnote{%
Swampland is defined as a space of EFTs that cannot be consistently coupled to quantum gravity.}
Such a tower of charged states have provided further motivations
to connect the WGC with the DC
\cite{Palti:2017elp,Grimm:2018ohb,Lee:2018urn,Lee:2018spm,Gendler:2020dfp}.

There are a number of swampland conjectures 
other than those mentioned above that have been proposed since \cite{Vafa:2005ui}
(see e.g. \cite{Palti:2019pca,vanBeest:2021lhn,Agmon:2022thq}).
Their \emph{plausibility} and \emph{usefulness} for constraining EFTs vary:
One of the major approaches for increasing the
\emph{plausibility} of a swampland conjecture is to 
realize many examples as low-energy limits of string theory
-- the leading candidate for the unified theory of all the forces in nature, including quantum gravity.
However, one drawback in this approach is that
to keep the theoretical calculations under control, 
most of the examples preserve supersymmetry, or remain just below the supersymmetry breaking scale.
Therefore, those examples from string theory provide limited support 
for the applicability of the swampland conjectures far below the string scale or the supersymmetry breaking scale.
Conversely,
for a swampland conjecture that is useful for constraining non-supersymmetric EFTs,
it is generically not very plausible that it originates in string theory.
This is the authors' view on the current status of the swampland programme.


In the meantime, there is one notable approach
for extending the \emph{usefulness} of swampland conjectures to lower energy scales:
assuming a swampland conjecture is satisfied at some high-energy scale,
trying to see if the swampland constraints propagate to lower energy.
In this regard,
Saraswat asked the following question on the WGC \cite{Saraswat:2016eaz}:
When the WGC is respected by an EFT at high energy,
can it be violated at low energy
via a spontaneous gauge symmetry breaking?
He studied this problem
in a $U(1) \times U(1)$ gauge theory spontaneously broken to $U(1)$
(we call this model \emph{Saraswat EFT}).
It has been shown that in the Saraswat EFT,
the most general version of
the WGC was not violated by the spontaneous gauge symmetry breaking
\cite{Furuuchi:2017upe}.\footnote{%
When the mass of a charged particle is below the UV cut-off scale of the EFT
bounded by \eqref{eq:mWGCbound},
the eWGC is automatically satisfied
\cite{Furuuchi:2017upe}.
Particles whose mass is above the UV cut-off scale of an EFT
are usually not included in the EFT.
But if we exclude a particle satisfying the eWGC \eqref{eq:eWGC}
that has mass above the UV cut-off,
the EFT can appear to violate the eWGC, 
but this should not lead to the inconsistency of the EFT.
It is possible to include a particle heavier than the UV cut-off in the EFT,
and for the above reason, Ref.~\cite{Furuuchi:2017upe} also studied the cases
where charged particles have masses above the UV cut-off scale.}


In the Saraswat EFT,
the Dirac monopole of the $U(1)$ Low Energy EFT
is composed of the Dirac monopoles of the $U(1)\times U(1)$
High Energy EFT,
connected by the magnetic flux tubes 
of the broken $U(1)$
(composite magnetic monopole, CMM).
The size of the CMM signifies the scale
at which the Low Energy EFT breaks down \cite{Furuuchi:2017upe}.

The Saraswat EFT provides an interesting example in which
above the UV cut-off scale of an EFT bounded by the mWGC \eqref{eq:mWGCbound},
another EFT takes over the description.
What we would like to test in this article is
whether
the mWGC$=$DC conjecture applies to the Low Energy EFT:
whether its UV description, 
the $U(1)\times U(1)$ High Energy EFT, 
has the tower of light states predicted by the DC.
In other words, we would like to examine
whether the WGC respected by the Low Energy EFT
implies the tower of light states 
in the High Energy EFT through the WGC$=$DC,
starting from the EFT that respects the WGC.
Indeed, the composite nature of the CMM
naturally gives rise to the tower of low-energy excitations.
We will examine 
the parameter dependence of the energy spectrum
of these low-energy excitations,
anticipating
that the large or small values of these parameters
correspond to the boundary of the moduli space of a UV theory,
as in eq.~\eqref{eq:e}.
It should be noticed that
while in the original DC,
the tower of light states that appears
as we approach the boundary of the moduli space
propagates in 4D space-time,
here, the low-energy excitations
are localized on the CMM.
However, as the localized energy contributes to the mass
when looked from a distance,
we think it natural to extend the DC to include
such localized excitations in the predicted tower
at the boundary of the moduli space.
In the following,
we examine this theme in depth.

\section{The Energy Spectrum of the CMM}\label{sec:composite}

\subsection{The Saraswat EFT}\label{sec:Saraswat}

The Saraswat EFT \cite{Saraswat:2016eaz}
has the gauge symmetry $U(1)_A \times U(1)_B$.
We denote the gauge couplings of the respective $U(1)$ groups as $e_A$ and $e_B$,
and take $e_A = e_B = e$ for simplicity. 
As the gauge fields in Abelian gauge theories
do not have a charge,
the normalization convention of the gauge coupling
cannot be based on the minimal self-coupling of the gauge fields.
We follow the convention that 
\emph{the charge of the smallest charge unit in the model is one}
\cite{Furuuchi:2017upe}.\footnote{%
The expression of the bound on the UV cut-off scale \eqref{eq:mWGCbound} 
depends on the normalization convention.}
We introduce a Higgs field $H$ with the charge vector $(Z,1)$, 
where the first number is the charge in the $U(1)_A$ gauge group
and the second number is the charge in the $U(1)_B$ gauge group.\footnote{%
As we accept the Dirac quantization condition,
the electric charges are quantized.} 
We also introduce matter fields
$\psi_A$ and $\psi_B$
whose charge vectors are given as
$(1,0)$ and $(0,1)$, respectively.
Only their charges and masses 
(we denote them $m_A$ and $m_B$)
are relevant in the following discussions.

The EFT description of the model is expected to
break down at some UV scale $\Lambda_{\rm UV}$
above which a new theory replaces the description.
In the Saraswat EFT,
this UV cut-off scale
is assumed to satisfy the bound from the magnetic WGC:
\begin{equation}
\Lambda_{\rm UV} \lesssim \frac{e M_P}{\sqrt{2}}\,.
\label{eq:mWGCboundH}
\end{equation}
Here, the factor $1/\sqrt{2}$ is a reminder that
the model has two $U(1)$ gauge groups
\cite{Cheung:2014vva},
though we are not concerned with such coefficients of order one in this article.

We will study a family of EFTs labelled by the Higgs charge $Z$
in $U(1)_A$ gauge group.
We are interested in the large $Z$ behaviours of the family of EFTs.
When taking $Z$ large, we fix $eZ \lesssim \Ord(1)$ 
so that the gauge coupling of the Higgs field remains
in the perturbative regime.

With a suitable symmetry breaking potential (see \cite{Furuuchi:2017upe} for an explicit example),
the Higgs field acquires a vacuum expectation value $v$
and the gauge symmetry $U(1)_A \times U(1)_B$
is spontaneously broken to $U(1)_{{\rm low}}$.
The charges with respect to the unbroken gauge group $U(1)_{{\rm low}}$
and those with respect to the broken gauge group $U(1)_{{\rm broken}}$
are listed in Table~\ref{table:chargelist}.
There,
the gauge coupling $e_{{\rm low}}$ of the $U(1)_{{\rm low}}$
is given as
\begin{equation}
e_{{\rm low}} := \frac{e}{\sqrt{1+Z^2}}\,.
\label{eq:elow}
\end{equation}
Note that the gauge coupling $e_{{\rm low}}$
follows our convention that the smallest charge unit becomes one:
as in Table~\ref{table:chargelist},
$\psi_A$ has the smallest charge magnitude with respect to the gauge group $U(1)_{{\rm low}}$.
We assume that 
the mass $\psi_A$ is smaller than the UV cut-off scale
$\Lambda_{\rm low}$
at which the low energy EFT description breaks down.
We assume that
the spontaneous symmetry breaking is described by the model.
This requirement leads to the condition $m_H$, $v$ $\ll \Lambda$.

\begin{figure}[htbp]
\centering
\includegraphics[width=5in]{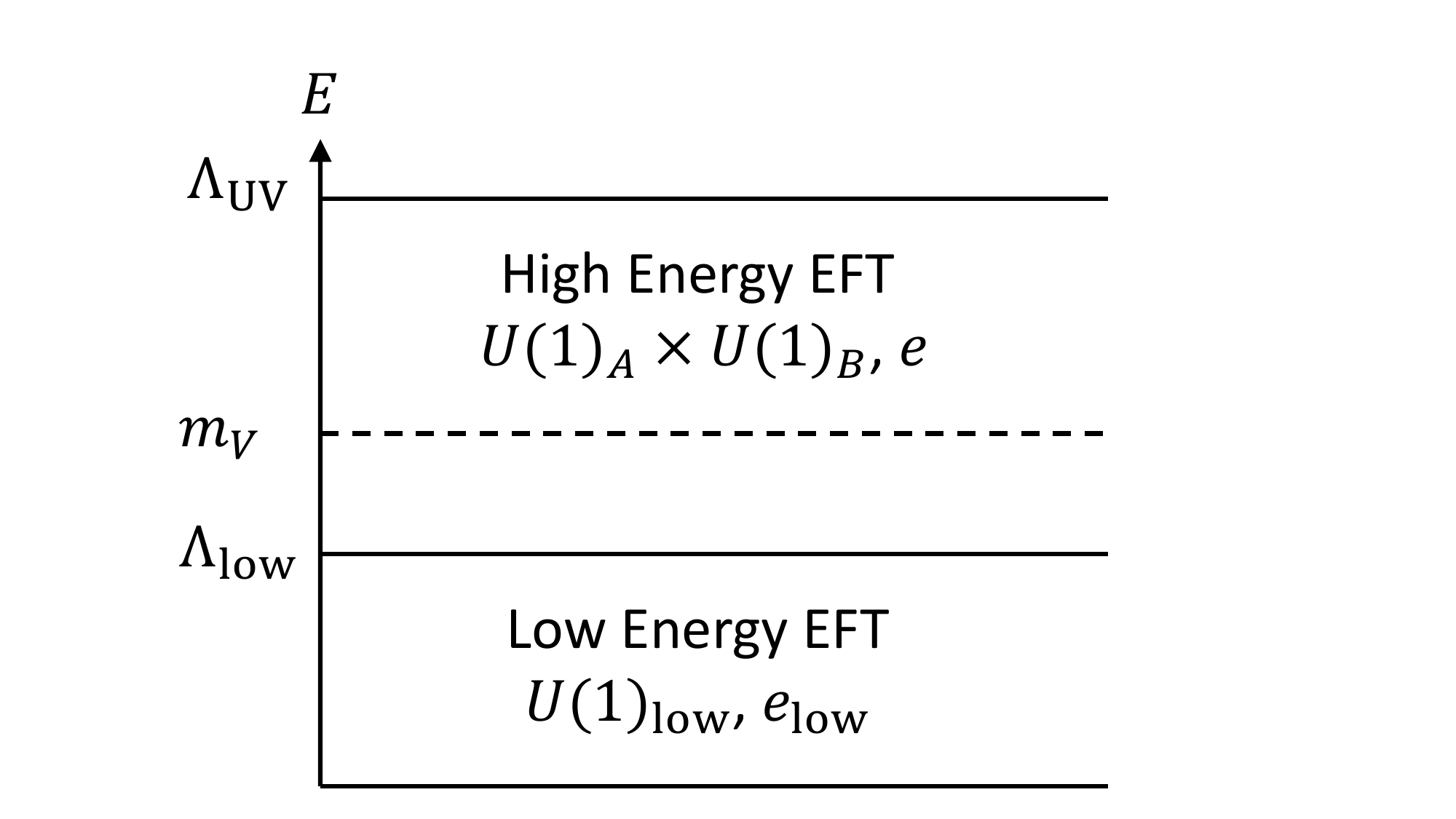} 
\caption{The High Energy EFT is the original Saraswat EFT.
It has the gauge group $U(1)_A\times U(1)_B$ with gauge coupling $e$.
Its UV cut-off scale is $\Lambda_{\rm UV}$.
The low energy EFT has the gauge group $U(1)_{{\rm low}}$ with gauge coupling $e_{{\rm low}}$.
Its UV cut-off scale is $\Lambda_{{\rm low}}$ is much lower than the spontaneous gauge symmetry breaking scale $m_V$
in the large $Z$ limit we consider.
\label{fig:EnergyScales}}
\end{figure}

The mass of the gauge field associated with the broken $U(1)$ gauge symmetry is given by
\begin{equation}
m_V = e v \sqrt{1+Z^2} \,.
\label{eq:gmass}
\end{equation}
Energy scales above $m_V$,
the gauge boson associated with 
the broken gauge group effectively recovers.
Below the spontaneous gauge symmetry breaking scale $m_V$,
effective descriptions of physics are given
by the EFT whose fields have masses smaller than $m_V$.
We call this EFT 
\emph{Low Energy EFT},
as opposed to the original full Saraswat EFT which we call 
\emph{High Energy EFT}.
However, as we explain below 
(originally described in \cite{Furuuchi:2017upe}),
the applicability of the Low Energy EFT turns out to be far below $m_V$
in the large $Z$ limit we consider.
In this article, 
we assume that $\psi_A$ and $\psi_B$ are included in the Low Energy EFT,
i.e. 
$m_A$, $m_B < \Lambda_{{\rm low}}$
(see \cite{Furuuchi:2017upe} for more general cases).
Strictly speaking, the coupling between
the charged fields and the Higgs field
can shift the masses of the charged fields.
What we assume here is that 
after these shifts in the masses,
the charged fields still have mass below $\Lambda_{\rm UV}$.\footnote{%
If we assume the mWGC bound \eqref{eq:mWGCbound},
and also the mass of the charge field is below the UV cut-off,
then, the eWGC is automatically satisfied \cite{Furuuchi:2017upe}.}
The relevant energy scales and the EFTs describing the physics
are schematically depicted in Fig.~\ref{fig:EnergyScales}.
Then, what the magnetic WGC predicts is that the 
Low Energy EFT with the $U(1)_{{\rm low}}$ gauge group
will break down below $\Lambda_{\rm low}$ bounded as 
\begin{equation}
    \Lambda_{{\rm low}} \lesssim e_{{\rm low}} M_P \,.
\label{eq:mWGCblow}
\end{equation}
To know what $\Lambda_{\rm low}$
should be,
we need to understand 
how the Dirac magnetic monopole
in the Low Energy EFT
is described in the High Energy EFT.
This is done in the next subsection.

\begin{table}
\begin{center}
\begin{tabular}{  | l | l | l | }
\hline
particle	& charge in $U(1)_A \times U(1)_B$ & 
charge in $U(1)_{{\rm broken}}\times U(1)_{{\rm low}}$ \\
\hline
$\psi_A$ & (e) $e (1,0)$ & (e) $e_{{\rm low}} (Z,-1)_{{\rm low}}$ \\
\hline
$\psi_B$ & (e) $e (0,1)$ & (e) $e_{{\rm low}} (1,Z)_{{\rm low}}$ \\
\hline
Higgs $H$ & (e) $e (Z,1)$ & (e) $e_{{\rm low}} (1+Z^2,0)_{{\rm low}}$ \\
\hline
$U(1)_A$ anti-monopole & (m) $\frac{2\pi}{e} (-1,0)$ & 
(m) $\frac{2\pi}{e_{{\rm low}}} 
\left(-\frac{Z}{1+Z^2},\frac{1}{1+Z^2}\right)_{{\rm low}}$ \\
\hline
$U(1)_B$ monopole & (m) $\frac{2\pi}{e} (0,1)$ & 
(m) $\frac{2\pi}{e_{{\rm low}}} 
\left(\frac{1}{1+Z^2}, \frac{Z}{1+Z^2} \right)_{{\rm low}}$\\
\hline
\end{tabular}
\caption{(e) and (m) indicate the electric- and magnetic charge,
respectively. 
We also indicated the electric- and magnetic gauge couplings to the charges.
\label{table:chargelist}}
\end{center}
\end{table}

\subsection{The CMM in the Saraswat EFT}\label{sec:CMM}

\begin{figure}[htbp]
\centering
\includegraphics[width=5in]{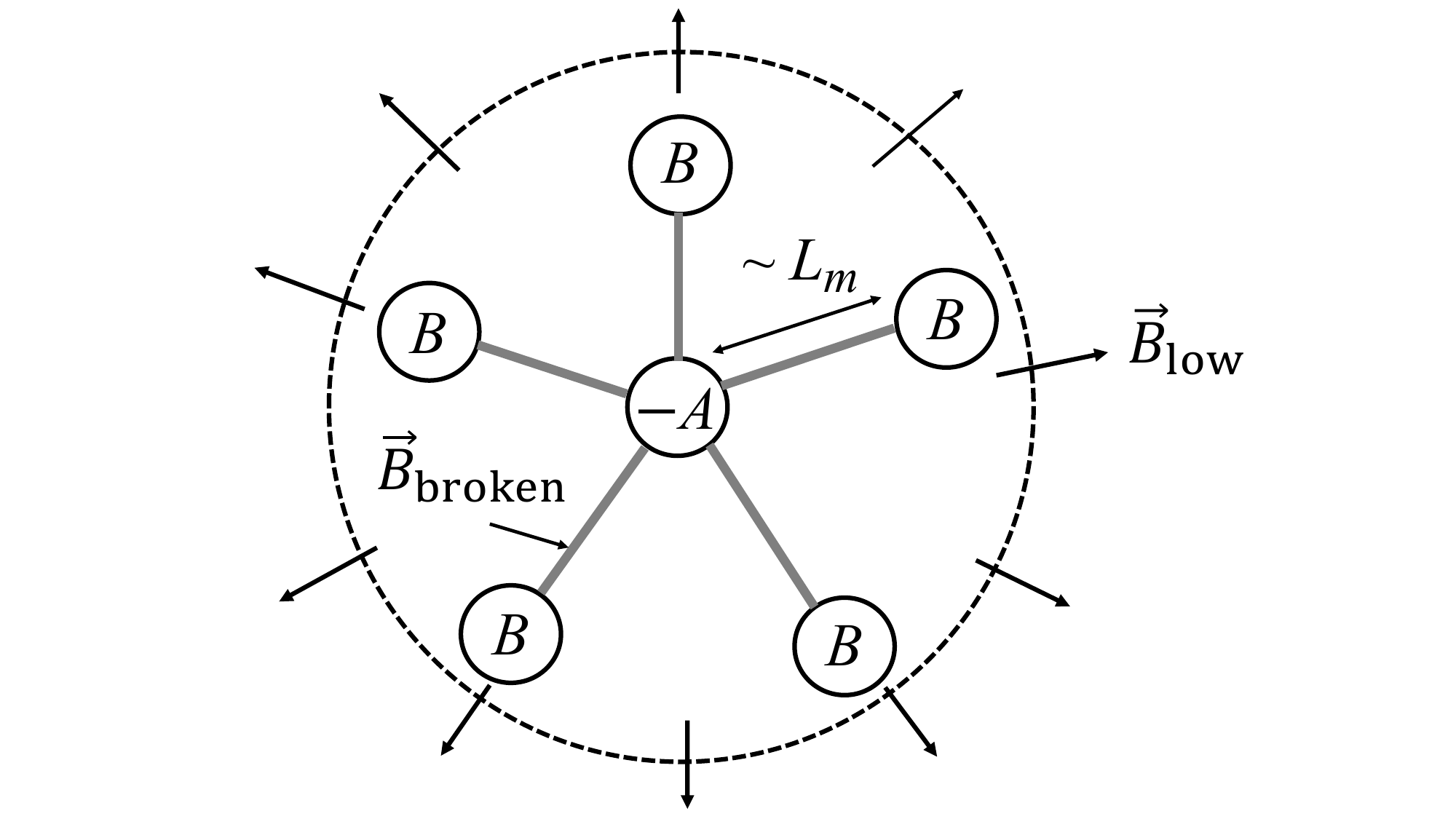} 
\caption{A schematic figure of a monopole
of $U(1)_{\rm low}$ with unit magnetic charge
(the magnetic flux of the
unbroken $U(1)_{\rm low}$
is indicated by $\vec{B}_{\rm low}$).
It consists of 
one anti-monopole of $U(1)_A$ 
(indicated by $-A$)
and
$Z$ monopoles of $U(1)_B$ (indicated by $B$),
connected by the Nielsen-Olesen flux tubes
with $U(1)_{{\rm broken}}$ magnetic flux
(indicated by $\vec{B}_{\rm broken}$) 
\cite{Saraswat:2016eaz}.
\label{fig:monopole}}
\end{figure}

As found in \cite{Saraswat:2016eaz},
the Dirac monopole with unit magnetic charge
with respect to $U(1)_{{\rm low}}$
is composed of
$Z$ Dirac monopoles of $U(1)_B$
and one anti-Dirac monopole of $U(1)_A$,
connected by Nielsen-Olesen flux tubes
\cite{Nielsen:1973cs,Nambu:1974zg}
(see Fig.~\ref{fig:monopole}).
Below, we will refer to the 
magnetic monopole of $U(1)_{\rm low}$ as 
\emph{composite magnetic monopole} (CMM),
whereas the magnetic monopoles of
$U(1)_{A}$ and $U(1)_{B}$ 
will be referred to as
\emph{constituent magnetic monopoles}.
The radius of the flux tube is of the order of
$1/m_V$,
and the tension is of the order of $v^2$.
Here, as in \cite{Saraswat:2016eaz},
we assume that the mass of the Higgs field is heavier
than the massive gauge boson,
in which case the symmetry-breaking vacuum 
acts like a type II superconductor.
This assumption is consistent 
with the assumptions we have made earlier,
$v$, $m_H$ $\ll \Lambda_{\rm UV}$ and $e Z \lesssim \Ord(1)$.
The size $L_m$ of the composite monopole
is estimated
from the balance between
the energy in the flux tubes
\begin{equation}
E_{\rm flux}(L) \sim Z v^2 L \,,
\label{eq:Eflux}
\end{equation}
and the magnetic repulsive potential energy between the monopoles due to the 
magnetic Coulomb force of the
unbroken gauge group $U(1)_{\rm low}$:
\begin{equation}
E_{\mathrm{MC}}(L) \sim Z^2 \left(\frac{2\pi}{e}\right)^2 \frac{1}{L} \,.
\label{eq:EMC}
\end{equation}
The forces balance at
\begin{equation}
L \sim L_m:= \frac{\sqrt{Z}}{e v}\, .
\label{eq:Lm}
\end{equation}
As explained in detail in \cite{Furuuchi:2017upe},
$L_m$ signifies the scale
at which the Low Energy EFT breaks down.
This is because 
the constituent magnetic monopoles
have fractional magnetic charges
with respect to the unbroken gauge group $U(1)_{\rm low}$,
see Table~\ref{table:chargelist}.
The fractional magnetic charges are inconsistent 
within the Low Energy EFT, 
hence the Low Energy EFT breaks down at the length scale $L_m$,
or the energy scale
\begin{equation}
\Lambda_{\rm low} := \frac{1}{L_m} \,.
\label{eq:LambdalowLm}
\end{equation}
It has been shown that the Low Energy EFT
respects the mWGC bound on the UV cut-off scale
if the High Energy EFT does \cite{Furuuchi:2017upe}:
\begin{equation}
 \Lambda_{\rm low} \lesssim e_{\rm low} M_P\,.
\label{eq:mWGCboundL}
\end{equation}
Note that the mass $M_{\rm CMM}$ of the CMM is dominated by the masses of the constituent magnetic monopoles:
\begin{equation}
M_{\rm CMM} \sim m_A + Z m_B \sim \frac{(1+Z) \Lambda_{\rm UV}}{e^2}\,.
\label{eq:MCMM}
\end{equation}
This is much bigger than the lower bound \eqref{eq:gmass}:
\begin{equation}
\frac{\Lambda_{\rm low}}{e_{\rm low}^2} \sim E_{\rm flux}(L_m) + E_{\mathrm{MC}}(L_m)\,.
\label{eq:MCMMb}
\end{equation}
Eq.~\eqref{eq:MCMMb} is what is expected for the mass of a magnetic monopole
if it were determined by the physics at the energy scale around $\Lambda_{\rm low}$.
However, the mass of the CMM is actually dominated by the masses of the constituent monopoles,
which are determined by the physics at the UV cut-off scale of the High Energy EFT.
The mWGC is still satisfied with $M_{\rm CMM}$ in eq.~\eqref{eq:MCMM}:
\begin{equation}
\frac{M_{\rm CMM}}{M_P}
\sim
\frac{(1+Z) \Lambda_{\rm UV}}{e^2 M_P}
\lesssim
\frac{(1+Z)}{e}
\sim \frac{2\pi}{e_{\rm low}}\,.
\label{eq:mWGCMCMM}
\end{equation}
In the inequality in eq.~\eqref{eq:mWGCMCMM}, 
we have used the model assumption \eqref{eq:mWGCboundH}
that
the High Energy EFT respects the mWGC.
Note that we are not concerned with coefficients of order one.

\subsection{A Proposal to Extend the DC to Include Localized Tower}\label{sec:extDC}

As explained in the introduction section,
the mWGC$=$DC conjecture
predicts that above the 
UV cut-off scale $\Lambda_{\rm low}$ of the Low Energy EFT
bounded by eq.~\eqref{eq:mWGCboundL},
a tower of light states appear.
The mass scale is predicted to
scale exponentially with a scalar field in a UV theory,
see eq.~\eqref{eq:DC}.
We are interested in the limit in which
$\Lambda_{\rm low}$ becomes small.
From eq.~\eqref{eq:LambdalowLm} and eq.~\eqref{eq:Lm},
this can be achieved by taking small $e$ or large $Z$.
However, when $Z$ is not large,
there's not much difference between the 
mWGC bound on the UV cut-off scale of the High Energy EFT and
that of the Low Energy EFT:
From the relation between the High Energy EFT gauge coupling $e$
and that of the Low Energy EFT \eqref{eq:elow},
the ratio of the bound on the UV cut-off scale of the High Energy EFT
\eqref{eq:mWGCboundH}
and that of the Low Energy EFT
\eqref{eq:mWGCboundL}
is given as
\begin{equation}
\frac{e M_P}{e_{\rm low} M_P} = \frac{e}{e_{\rm low}} = \sqrt{1+Z^2} \,.
\label{eq:hie1}
\end{equation}
If there is not much difference between $\Lambda_{\rm UV}$
and $\Lambda_{\rm low}$,
the High Energy EFT may not be able to describe the
tower of light states
that appears above $\Lambda_{\rm low}$:
We would need a UV theory of the High Energy EFT to describe it.
We are interested in the tower of light states described by the High Energy EFT,
so we will consider the limit $Z \gg 1$.
We assume that the large $Z$ limit
is related to the DC in a UV theory:
\begin{equation}
\frac{m}{m_0} \sim Z^{-\beta}\,.  
\label{eq:Z}
\end{equation}
Here, $\beta$ is a positive constant of order one, 
and $m$ is the mass scale of the tower predicted by the DC, and $m_0$ is a constant, see \eqref{eq:DC}.
A similar assumption was made in 
the Dark Dimension scenario for the positive cosmological constant \cite{Montero:2022prj},
extrapolating from the string theory examples 
in the case of the negative cosmological constant \cite{Lust:2019zwm}.
The assumption \eqref{eq:Z} cannot be addressed within the framework of EFT.
For that, one has to resort to the UV complete theory of quantum gravity i.e. string theory.
Our aim here is to explore EFT examples in which
a tower of light states appear at the UV cut-off scale of the EFT bounded by mWGC \eqref{eq:mWGCblow}.
Such a tower of light states 
then provide EFT supports for 
the WGC$=$DC connection through the identification \eqref{eq:Z}.
Later on, we extend the DC to include localized low-energy excitations
(see \eqref{eq:extDC} below).
Then, the mass scale $m$ of the tower will be replaced
with the energy excitation scale $E$, see \eqref{eq:ZE} below.

While taking $Z\gg 1$, however,
we need to keep $eZ \lesssim \Ord(1)$ fixed
in order for the High Energy EFT 
to remain perturbative, 
including the minimal gauge coupling to the Higgs field.
Therefore, we will consider the limit
$e \ll 1$, $Z \gg 1$,
with $eZ \lesssim \Ord(1)$ fixed.

While the mWGC bound on $\Lambda_{\rm UV}$ \eqref{eq:mWGCboundH} scales with $e$,
we do not scale $\Lambda_{\rm UV}$ itself:
For practical EFT model-building purposes,
we don't expect to take $Z$ too large, 
say $Z \sim \Ord(10)$ to at most $Z \sim \Ord(100)$.
In the limit $eZ \lesssim \Ord(1)$ fixed,
this amounts to taking
$e \lesssim \Ord(10^{-1}) - \Ord(10^{-2})$.
The mWGC bound on $\Lambda_{\rm UV}$ \eqref{eq:mWGCboundH}
still accommodates a large range for the new particle physics models beyond the Standard Model.
Not scaling $\Lambda_{\rm UV}$ with $e$
also guarantees that the tower of low-energy excitations
we are studying is associated with the breakdown of the Low Energy EFT,
not that of the High Energy EFT.
We also keep $v/\Lambda_{\rm UV} \ll 1$ fixed while scaling $e$.
Note that in this limit,
the hierarchy between the UV cut-offs of the High Energy EFT and the Low Energy EFT becomes
\begin{equation}
\frac{\Lambda_{\rm UV}}{\Lambda_{\rm low}}
=
\frac{\sqrt{Z}}{ev} \Lambda_{\rm UV}
=
\frac{Z^{3/2}}{(eZ)}\frac{\Lambda_{\rm UV}}{v}\,,
\label{eq:hie2}
\end{equation}
refining the considerations around eq.~\eqref{eq:hie1}.

What would be the tower of light states that appears
above $\Lambda_{\rm low}$ and that is described by High Energy EFT?
It is natural to suspect that
the tower of new degrees of freedom
is associated with the CMM,
considering the central role the CMM plays
in formulating the mWGC.
In fact, 
due to the composite nature of the CMM,
it is expected to have a tower of excitated states.
Below in Sec.~\ref{sec:excitations},
we will verify that the energy scale
of these excitations
scales 
as the DC predicts.
However, the original DC predicts that
the tower of light states propagating in 4D space-time,
while here, the low-energy excitations
are localized on the CMM:
The CMM is so heavy (see \eqref{eq:MCMM}) that
the momentum transfers even in High Energy EFT
do not alter the momentum of the CMM significantly.
Thus we propose to extend
the DC to include the tower of localized low-energy excitations
that appear as we approach the boundary of the moduli space:
Our extended DC proposes that 
\emph{as a scalar field approaches the boundary of the moduli space,
a tower of states whose energy scale $E$ 
decreases as below appears:
\begin{equation}
E \sim E_0 \exp \left[ - \beta' \frac{\phi}{M_P} \right] \,.
\label{eq:extDC}
\end{equation}
Here, $\beta'$ is a constant of order one with $\beta' \phi > 0$,
and $E_0$ is a constant energy scale.}
As the energy contributes to the mass,
this is a natural extension of the DC.
Considering the central role of the magnetic monopole in formulating the mWGC bound \eqref{eq:mWGCbound},
we expect that such a tower localized on the magnetic monopole
is sufficient for eliminating the inconsistency with quantum gravity
relevant to the WGC and the WGC$=$DC conjecture.

In passing, following the extended DC conjecture,
we modify \eqref{eq:Z} to
\begin{equation}
\frac{E}{E_0} \sim Z^{-\beta}\,.  
\label{eq:ZE}
\end{equation}

Below, we examine whether the energy scale of the excitations of the CMM
agrees with the prediction of the DC extended to include the localized tower \eqref{eq:ZE}.

\subsection{Excitation Energy Spectrum of the CMM}\label{sec:excitations}

Armed with the understanding of the CMM in the Saraswat EFT 
from the previous subsections,
we now analyze the energy spectrum of the CMM.
We are more interested in the lower energy excitations
in the limit $e \ll 1$, $Z \gg 1$, with $eZ \lesssim \Ord(1)$ fixed.
It turns out that the rotational modes
are the lowest energy excitations in this limit.
These are discussed in Sec.~\ref{sec:rot}.
We also study the oscillatory modes in Sec.~\ref{sec:osc}
and compare the energy scales with the rotational modes.\footnote{%
Here, we are not much concerned with numerical factors of order one.
More precise estimates of the energy spectrum of the CMM,
modelling the configuration of the 
large number of the monopoles in $U(1)_B$ gauge group
as a spherical shell, 
are given in \cite{MoreshwarTh} 
(a soft copy available upon reasonable request).}

\subsubsection{Rotational Modes}\label{sec:rot}

The Hamiltonian for the rotational modes
of the CMM is given by
\begin{equation}
H_{\rm rot}
\sim
\frac{J^2}{2 Zm_B L_m^2}\,,
\label{eq:Hrot}
\end{equation}
where $J^2$ is the total angular momentum square, and
$m_B$ is the mass of the Dirac magnetic monopole in $U(1)_B$ gauge group
estimated as in eq.~\eqref{eq:monopolemass}:
\begin{equation}
m_B \sim \frac{\Lambda_{\rm UV}}{e^2}\,.
\label{eq:mB}
\end{equation}
In quantum mechanics, the eigenvalues of $J^2$ are quantized 
as $j(j+1)$, where $j$ is a positive integer.
Here, we treat the CMM as a rigid body.
The justification of it is given shortly.
Putting \eqref{eq:mB} 
and \eqref{eq:Lm} into \eqref{eq:Hrot}, 
the energy eigenvalues are estimated to be
\begin{equation}
E_{{\rm rot},j}
\sim
\frac{e^4}{2Z^2} v \left(\frac{v}{\Lambda_{\rm UV}}\right) j(j+1)
\sim
\frac{(eZ)^4}{2Z^6} v \left(\frac{v}{\Lambda_{\rm UV}}\right)j(j+1) \,.
\label{eq:Erotj}
\end{equation}
In the limit of our interest, i.e. 
$e \ll 1$, $Z \gg 1$,
with $eZ$ fixed,
the energy scale of the rotational excitations
scales as $Z^{-6}$.
Anticipating that our large $Z$ limit 
is related to the extended DC
as in eq.~\eqref{eq:ZE},
the result \eqref{eq:Erotj} is in accordance with the prediction of the extended DC:

It is useful to compare the energy scale of the rotational modes
with the UV cut-off scale $\Lambda_{\rm low} \sim 1/L_m$ of the Low Energy EFT:
\begin{equation}
E_{{\rm rot},j}
\sim
\Lambda_{\rm low}
\frac{(eZ)}{Z^{5/2}}
\left( \frac{v}{\Lambda_{\rm UV}} \right)\,.
\end{equation}
We observe that the energy for rotating 
the CMM is well within the energy scale of the Low Energy EFT.
In other words,
the Low Energy EFT degrees of freedom
can excite the rotational modes of the CMM, provided that the CMM exists.
However, this accessibility of the rotational excitations of the CMM to the Low Energy EFT
does not mean the breakdown of the Low Energy EFT:
Heavy particles whose masses are above the UV cut-off of an EFT
can be incorporated in the EFT,
like, for example, the heavy quark effective theory \cite{Manohar:2000dt}.
Different excitations of the CMM
can be incorporated into the Low Energy EFT as different particles.
The masses and the interactions of the excited CMM with the Low Energy EFT degrees of freedom
are inputs from the High Energy EFT to the Low Energy EFT:
The Low Energy EFT cannot determine them.

One might be a bit puzzled 
in view of the WGC$=$DC conjecture
that the energy scale of the excitations
is not given by $\Lambda_{\rm low}$
but has additional powers of $Z$.
However, from our assumption \eqref{eq:ZE},
the difference in the powers of $Z$
is translated to the difference of the 
order one coefficients ($\beta$ in eq.~\eqref{eq:ZE}).
As the DC does not precisely specify the order one constant,
we think this difference is not against the WGC$=$DC conjecture.

Below, we check that the 
distortion of the CMM due to the centrifugal force
is negligible in the limit we are considering.
The centrifugal force is estimated as
\begin{equation}
F_{\rm cent}
\sim \frac{J^2}{Z m_B L_m^3}\,,
\label{eq:Fcent}
\end{equation}
is much smaller than the forces from the flux tube tension or the magnetic Coulomb around $L \sim L_m$:
The ratio of the centrifugal force and the flux tube force is estimated as:
\begin{equation}
\frac{F_{\rm cent}}{F_{\rm flux}}
\sim
\frac{J^2}{Z v^2 \cdot Z m_B L_m^3}
\sim 
J^2 \frac{e^5}{Z^{7/2}} \frac{v}{\Lambda_{\rm UV}}
\sim
J^2 \frac{(eZ)^5}{Z^{17/2}} \frac{v}{\Lambda_{\rm UV}}
\,,
\label{eq:rFfFT}
\end{equation}
This ratio is extremely small
in the limit under consideration,
$e \ll 1$, $Z \gg 1$, with $eZ \lesssim \Ord(1)$ fixed.
In other words,
our rigid body approximation is good until
$j$ reaches the critical value $j_c \sim Z^{17/4} \sqrt{\Lambda_{\rm UV}/v}$.

In passing, we note that
the rotation of a single constituent monopole in $U(1)_B$ gauge group
costs more energy
due to the smaller effective mass
(compare with eq.~\eqref{eq:Hrot}).

\subsubsection{Oscillatory Modes}\label{sec:osc}

The Hamiltonian for the overall size $L$ of the CMM is given as
\begin{equation}
H_{\rm osc}
\sim
\frac{Z m_B}{2} \dot{L}^2 + E_{\rm flux}(L) + E_{\mathrm{MC}}(L) \,,
\label{eq:Hosc}
\end{equation}
where $E_{\rm flux}$ and $E_{\mathrm{MC}}$ are given in \eqref{eq:Eflux} and \eqref{eq:EMC},
respectively.
\eqref{eq:Hosc} can be rewritten as
\begin{equation}
H_{\rm osc}
\sim
\frac{Z m_B}{2} \dot{L}^2 + V(L) \,,
\label{eq:Hosc2}
\end{equation}
where
\begin{equation}
V(L) = K \left( \frac{L_m^3}{L} + L L_m\right)\,,
\label{eq:VL}
\end{equation}
and
\begin{equation}
K = \frac{Z^2}{L_m^3} \left( \frac{2\pi}{e} \right)^2\,.
\end{equation}
We approximate the 
potential around $L=L_M$
up to the quadratic order in the Taylor expansion
(the justification to be given shortly below)
to obtain\footnote{%
We are interested only in the difference between the energy of the excited states and that of the ground state.}
\begin{equation}
E_{\mathrm{osc},\,n}
\sim
\omega n\,, 
\end{equation}
where
\begin{equation}
\omega 
= \sqrt{\frac{K}{Zm_B}}
= \frac{e^{3/2}}{Z^{1/4}} (2\pi)  v \sqrt{\frac{v}{\Lambda_{\rm UV}}} 
=  \frac{(eZ)^{3/2}}{Z^{7/4}} (2\pi) v \sqrt{\frac{v}{\Lambda_{\rm UV}}} \,.
\label{eq:omega}
\end{equation}
In the limit we consider, $e \ll 1$, $Z \gg 1$, with $eZ \lesssim$ fixed,
the energy scale $\omega$ of the oscillatory modes 
scales as $\omega \propto Z^{-7/4}$,
thus the parameter dependence of the
energy scale of the 
oscillatory modes
fulfills the prediction of the WGC$=$DC conjecture,
see \eqref{eq:ZE}.
Note that the energy scale of 
the rotational modes decreases faster than the oscillatory modes
by a factor of $Z^{-17/4} \sqrt{v/\Lambda_{\rm UV}}$.

Again, it is useful to compare the energy scale of the oscillatory modes
with the UV cut-off scale $\Lambda_{\rm low} = 1/L_m$ of the Low Energy EFT.
We obtain
\begin{equation}
\omega \sim
\Lambda_{\rm low}
\frac{(eZ)^{1/2}}{Z^{1/4}}
(2\pi)
\sqrt{\frac{v}{\Lambda_{\rm UV}}}\,.
\end{equation}
Again, we observe that
the oscillatory modes of the CMM can be excited
by the Low Energy EFT degrees of freedom.
As explained for the rotational modes,
this does not mean the breakdown of the Low Energy EFT.

From \eqref{eq:Hosc2}, 
we observe that approximating the potential by the quadratic potential 
is valid up to $|L-L_m| \sim L_m$.
By setting
\begin{equation}
\omega n_c 
\sim V(2L_m)\,,
\end{equation}
we obtain the critical excitation level $n_c$ around which
the harmonic oscillator approximation breaks down.
We have
\begin{equation}
n_c \sim
\frac{KL_m^2}{\omega}
\sim
Z^{17/4}
(eZ)^{-5/2}
\sqrt{\frac{\Lambda_{\rm UV}}{v}} \gg 1 \,,
\end{equation}
i.e. in the current limit,
the approximation by the harmonic oscillator is valid up to a very large excitation level $n_c$.

In passing, we note that
oscillations of a single constituent monopole in $U(1)_B$ gauge group
costs more energy than the collective oscillations,
due to the smaller effective mass (compare with \eqref{eq:omega}).

\subsection{Remarks on a Tower 
Propagating in Space-Time}\label{sec:closed}

In perturbative string theory,
the existence of open strings necessitates the existence of closed strings.
Open strings can be constrained to move along a D-brane,
but closed string propagate in the whole space-time.
Similarly, while the end points of the ``open'' magnetic flux tubes 
are attached to the heavy constituent magnetic monopoles
and cannot move freely in 4D space-time,
closed magnetic flux tubes can propagate freely in 4D space-time.
Their excitation energy scale is determined by the tension of the magnetic flux tubes,
whose magnetic flux is quantized in accord with the charge of the Higgs field
in the broken $U(1)_{\mathrm{broken}}$ gauge group:
\begin{equation}
\omega_{\text{closed, osc}} \simeq v \,.
\label{eq:closedosc}
\end{equation}
In the large $Z$ limit with fixed $eZ$ we are considering,
this energy scale is parametrically larger
than the inverse of the CMM size scale \eqref{eq:Lm}:
\begin{equation}
\Lambda_{\mathrm{low}}:= \frac{1}{L_m} \simeq \frac{e v}{\sqrt{Z}} = \frac{(eZ)v}{Z^{3/2}}
\ll v \simeq \omega_{\text{closed, osc}}\,.
\label{eq:closedhie}
\end{equation}
Note that the radius of the flux tube is of the order of $1/m_V$,
where $m_V$ is the mass of the gauge field associated with the 
broken $U(1)$ gauge group \eqref{eq:gmass}.
This is of the same order as $v$ in the current large $Z$ limit with $eZ$ fixed.
Therefore, magnetic flux tubes above several excitation energy levels
can effectively be described as closed strings, i.e., with Nambu-Goto action.
What eq.~\eqref{eq:closedhie} means is that the space-time propagating tower
is not directly associated to the breakdown
of the Low Energy EFT.
But the space-time propagating tower does exist
in the High Energy EFT.
Qualitatively speaking,
the large number ($\sim Z$) of magnetic charges in the broken $U(1)$ gauge group
stretch the magnetic flux tubes by the magnetic Coulomb interactions
to make the CMM parametrically larger than
the inverse of the energy scale of the excitations of the closed magnetic flux tube.
%
As we have seen in \eqref{eq:Erotj} and \eqref{eq:omega},
the excitation energy scales of the CMM are also
parametrically smaller than the excitation energy scale of 
the closed magnetic flux tube \eqref{eq:closedosc}.

\section{Summary and Discussions}\label{sec:Discussions}

In this article,
we studied the low-energy excitations 
of the composite magnetic monopole (CMM)
in the Sawraswat EFT.
Our aim was to examine that,
assuming swampland conjectures are respected in a UV theory,
whether their constraints descend
to lower energy scales.
This is important
for swampland conjectures 
to be useful in constraining
EFTs far below the quantum gravitational scale or the string scale,
or the supersymmetry breaking scale above which 
calculations for a string compactification may be under control.
We examined the connection between the weak gravity conjecture (WGC)
and the distance conjecture (DC).
However, unlike the WGC whose statement can be tested
within the framework of EFT (apart from the subtlety regarding
stable heavy particles whose masses are above the UV cut-off scale of the EFT
\cite{Furuuchi:2017upe}),
here,
the scalar field relevant to the DC
and the tower of light states 
are both beyond the validity of the Low Energy EFT.
For the former point, we study the behavior of the EFT
in the small ($e$)/large ($Z$) limit of its parameters.
Their relation to the scalar field responsible for the DC was well-motivated but an assumption,
which should be addressed in the UV complete theory of quantum gravity, namely, string theory.
For the latter point,
while the tower of light states is signaling the breakdown of the Low Energy EFT,
it can still be described within the High Energy EFT.
We showed that the low-energy excitations of the CMM were broadly
in accordance with the prediction of the DC.
However,
while the original DC predicts a tower of light states 
propagating in 4D space-time at the boundary of the moduli space,
here, the low-energy excitations are localized around the CMM.
Thus, we were led to extend the DC to
include the tower of low-energy excitations localized in space.
Considering the central role played by the magnetic monopole
in formulating the magnetic weak gravity conjecture (mWGC) bound \eqref{eq:mWGCbound},
we expect that such a tower localized on the magnetic monopole
would be sufficient for eliminating the inconsistency with the quantum gravitational effects
relevant to the WGC and the WGC$=$DC conjecture.
However, if one does not accept our proposal to extend the DC
including localized low-energy excitations,
the model still offers a possible route to break 
the connection of the WGC to the original DC at low energy,
even if there's one at high energy.
Either way, we demonstrated that 
the Saraswat EFT provides an important testing ground 
for whether and precisely how the constraints from swampland conjectures 
descend (partially, or not at all) to lower energy scales.

It has been proposed that
the kinetic terms gauge fields
emerge from the quantum gravitational scale
by integrating out towers of charged particles
(emergence proposal)
\cite{Heidenreich:2017sim}.
It was further proposed that the bound on the UV cut-off from mWGC \eqref{eq:mWGCbound}
derives from the quantum gravitational scale in this proposal.
For this scenario to work,
the towers of charged particles must propagate in 4D space-time.
This appears to be in sharp contrast to the
localized tower proposed to be relevant to the mWGC bound in this article.
However, 
we emphasize that our interest is assuming that swampland conjectures
are respected in a UV theory,
whether and how their constraints descend to lower energy EFTs.
As emphasized in \cite{Furuuchi:2017upe} and reviewed in this article,
the Low Energy EFT breaks down at
the size of the CMM, without directly resorting to quantum gravitational effects.
Our view is that while the emergence proposal of \cite{Heidenreich:2017sim}
may explain the bound on the UV cut-off scale in mWGC in the highest energy EFT,
it may not directly explain the bound on the UV cut-off scale in mWGC at lower-energy EFTs.
As discussed in Sec.~\ref{sec:closed}, it should be noted that 
a tower of states propagating in 4D space-time
does arise from closed magnetic flux tubes
in the High Energy EFT, its excitation energy scale
is parametrically higher than the breakdown energy scale 
$\Lambda_{\mathrm{low}}$ of the Low Energy EFT.
Hence the tower from the excitations of the closed magnetic flux tubes
is not directly relevant to the breakdown of the Low Energy EFT.

Relatedly,
the DC has a more restrictive version called
the emergent string conjecture (ESC),
which proposes that the lightest tower of light states predicted
by the DC 
is either Kaluza-Klein (KK) modes or excitations of weakly coupled fundamental strings
\cite{Lee:2019wij,Lee:2019xtm}.
If the ESC applies to an EFT, 
after the breakdown of the EFT,
there is no EFT description in the original space-time dimensions.
In contrast,
in our extended DC that accommodates
local low-energy excitations,
after the breakdown of the EFT,
another EFT in the same space-time dimensions can take over.
The Sawaswat model provides an explicit example:
Above the energy scale at which the Low Energy EFT breaks down,
the High Energy EFT takes over.
Our extended DC proposal
is more in harmony 
with the traditional philosophy of EFT:
A breakdown of an EFT does not necessarily
signal the appearance of an exotic quantum gravity theory or string theory:
It merely requires another EFT that can describe higher energy scales.
This has an advantage in enlarging the applicability of the EFT framework,
see, for example, \cite{Furuuchi:2017upe} for the application of the 5D version of the Saraswat EFT
to achieve large-field inflation.
On the other hand,
the Saraswat EFT, as an EFT,
cannot prove or disprove swampland conjectures;
this was not the goal of the current study.
We started with an assumption that the High Energy EFT
respects the mWGC bound \eqref{eq:mWGCboundH},
and studied whether the EFT
follows the prediction of the WGC$=$DC conjecture,
with an additional assumption that the large $Z$ limit is related to the DC, as in \eqref{eq:ZE}.
Our largely affirmative results would motivate embedding the Saraswat EFT into string theory.
The Saraswat EFT appears as a fairly innocent EFT, apart from the large $Z$.
If the Saraswat EFT
can be realized as a low-energy limit of string theory, i.e.
if it is not in the swampland,
one possibility is that
swampland conjectures, like the ESC,
do not constrain EFTs at the energy scales far below the 
quantum gravitational scale or string scale.
Examining this possibility was indeed the original motivation of the Saraswat EFT \cite{Saraswat:2016eaz}.
In the meantime,
it is widely believed
that the resolution of the black hole information paradox
requires quantum gravitational non-locality (see, for example, \cite{Giddings:2009ae,Almheiri:2020cfm}).
There is another possibility that such quantum gravitational non-local effects
might be at work to constrain EFTs even far below the Planck scale.
However, until we precisely understand 
how quantum gravitational non-locality works
for a particular swampland conjecture,
this possibility just remains as a possibility
(see, however, \cite{Aalsma:2019ryi,Aoufia:2026bau} for an investigation of 
the role of stringy non-locality in the WGC).
We should note that 
the proposed derivations of the WGC based on
low-energy arguments 
have so far failed 
(see, for example, Sec.~6 of the review \cite{Harlow:2022ich}).
There is a possibility of a more gradual difference in applicability:
more restricted versions of the swampland conjectures, like the ESC,
may be less applicable at lower energy scales.
A related theme in the lattice WGC 
has recently been investigated in \cite{Etheredge:2025rkn}.
In this regard, our extended DC proposal is even broader than the original DC, rather than restrictive,
including the possibility of a localized tower of low-energy excitations.
Therefore, it might have a greater chance of being applicable to EFTs 
at lower energy scales. 
The theme of the current work,
how a swampland conjecture propagates to lower energy scales,
has to be studied for the respective swampland conjectures
if such a gradual difference in applicability applies.

Including the localized tower of low-energy excitations
may also be motivated by D-branes in string theory,
as open strings are localized on D-branes 
\cite{Dai:1989ua,Polchinski:1995mt}.
This motivation may be enhanced further by the ESC,
in the case where the tower is from weakly coupled fundamental strings:
The open fundamental strings can be constrained on D-branes localized in space. 
It will be interesting to look for such localized towers
in the infinite distance limits in string theory.
For fundamental strings, usually, the energy scale of the open string mass spectrum
and that of the closed string
are both determined by the string tension and at the same order.
Here, the energy scale of the excitations of the CMM
is parametrically smaller than the 
excitation energy scale of the closed magnetic flux tube
due to its composite nature,
as discussed in Sec.~\ref{sec:closed}.
Therefore, we concluded that
the tower of excitations of the closed magnetic flux tube
was not directly relevant to the breakdown of the Low Energy EFT.
However, both open and closed magnetic flux tubes have the same tension.
Through this relation, it is possible that
there is an intricate consistency relation 
between the appearance of the tower of closed magnetic flux tube excitations
and the breakdown of the Low Energy EFT, which is not obvious at this moment.
We would like to point out that
constructing examples in string theory compactifications
is a promising approach to support for swampland conjectures,
but a drawback is that
controlled calculations have 
only been done in the backgrounds that preserve supersymmetry.
If some parts of the swampland conjectures
are to survive far below the supersymmetry breaking scale,
we think it quite possible that
the roles of fundamental strings in the ESC
are replaced by effective strings, 
like the magnetic flux tubes studied in this article.
This is along with our earlier suggestion that
less restrictive swampland conjectures may tend to apply at lower energy scales.


As mentioned above, one of the major approaches
to test swampland conjectures
is to construct (counter) examples in string theory. 
In this context, it may be worthwhile to recall that
magnetic monopoles
can be realized as
the end-points of D1-branes on D3-branes
\cite{Diaconescu:1996rk}.
It will be interesting to embed the Saraswat EFT in string theory,
identify the moduli field responsible for the DC applied to a large $Z$ \eqref{eq:Z},
and clarify the stringy origin of the CMM and the possible relation to the ESC.
It should be noted that 
while the charge
$Z$ may appear as a large charge,
it is so in the unit of the smallest charge:
The gauge coupling of the Higgs field was kept in perturbative regime, $eZ \lesssim \Ord(1)$.
If we regard the gauge coupling of the Higgs field as a unit instead,
large $Z$ limit is the small charge limit of the charged field 
($\psi_A$ in Table~\ref{table:chargelist}).
For this reason,
the construction of milli-charged particles in string theory \cite{Shiu:2013wxa}
may be useful in 
embedding the Saraswat EFT with a large $Z$ in string theory.\footnote{%
The milli-charges studied in \cite{Shiu:2013wxa} are generically irrational.
This would not pose a problem in constructing a Saraswat-like model
if the problem were solved in string theory.} 
If we can realize a model closely related to the Saraswat EFT in string theory,
we should also check if the limit, $e \ll 1$, $Z \gg 1$, with $eZ \lesssim \Ord(1)$ fixed,
leads to decompactification,
another possibility in the ESC,
before the localized tower we found appears.

\vspace*{4mm}
\begin{center}
\textbf{Acknowledgments}
\end{center}
\vspace*{-1.5mm}
We would like to thank 
Koushik Dutta and Rashmikanta Mantry for useful discussions.
This work was supported in part by the project file no.~DST/INT/JSPS/P-344/2022.
We thank the participating members and the supporting staff, in particular, 
the JSPS side PI Keisuke Izumi for various arrangements.
KF thanks Toshifumi Noumi for 
the hospitality during his visit to the University of Tokyo, Komaba, 
supported by the above project,
and for useful discussions.
Manipal Centre for Natural Sciences, \textsl{Centre of Excellence}, 
Manipal Academy of Higher Education (MAHE) 
is acknowledged for facilities and support.

\bibliography{mWGCDC}
\bibliographystyle{utphys}
\end{document}